# First-principles study of TMNa$_n$ (TM= Cr, Mn, Fe, Co, Ni; $n$ = 4-7) clusters.

Kalpataru Pradhan,[1] Prasenjit Sen,[1*] J. Ulises Reveles,[2] and Shiv N. Khanna[2]

[1]*Harish-Chandra Research Institute, Chhatnag Road, Jhunsi, Allahabad 211019, INDIA.*

[2]*Department of Physics, Virginia Commonwealth University, Richmond VA, 23284-2000, USA.*

*Corresponding author Fax: +91-532-2567748, E-mail: prasen@hri.res.in

## Abstract

Geometry, electronic structure, and magnetic properties of TMNa$_n$ (TM=Cr-Ni; $n$ = 4-7) clusters are studied within a gradient corrected density functional theory (DFT) framework. Two complementary approaches, the first adapted to all-electron calculations on free clusters, and the second been on plane wave projector augmented wave (PAW) method within a supercell approach are used. Except for NiNa$_n$, the clusters in this series are found to retain the atomic moments of the TM atoms, and the magnetic moment presented an odd-even oscillation with respect to the number of Na atoms. The origin of these odd-even oscillations is explained from the nature of chemical bonding in these clusters. Differences and similarities between the chemical bonding and the magnetic properties of these clusters and the TMNa$_n$ (TM = Sc, V and Ti; $n = 4 – 6$) clusters on one hand, and TM-doped Au and Ag clusters on the other hand, are discussed.

## 1. Introduction

Extensive research combining experiments on size selected clusters and the companion theoretical investigations, over the past two decades, has shown that the reduction in size can lead to qualitative changes in the magnetic behavior.[1-3] For example, small clusters of itinerant ferromagnetic elements like Fe, Co, and Ni are found to display super-paramagnetic relaxations[4] while the clusters of non-magnetic solids like Rh are found to be magnetic.[5,6] The magnetic moments of Fe$_n$, Co$_n$, and Ni$_n$ clusters containing two to a few dozen atoms are found to be almost 30% larger than the bulk solids leading to the optimism of designing stronger magnets.[7,8] Clusters of Mn$_n$, that exhibits complex ferromagnetic order in the bulk, display equally fascinating behavior.[9-12] The current thinking is that while Mn$_2$ has antiferromagnetically ordered localized moments, Mn$_3$ and Mn$_4$ have ferromagnetically aligned moments. The spin canting starts in Mn$_5$ and Mn$_6$, eventually evolving towards the bulk complex order.[9] While these are exciting findings, the practical applications of clusters require processes where the clusters could maintain the novelty exhibited in the free cluster. One of the suggested routes is to deposit clusters on substrates and this has generated considerable interest in studying the effect of substrates on the deposited species.[13-15] In order to maintain the properties of the free system, the logical choice is then to explore substrates that are chemically inert. This is because the deposition on substrates favoring stronger interaction is expected to quench the magnetic moment of the deposited species due to enhanced bonding that favors the filling of orbitals with electrons of opposite spins, thus negating the large moments attained in the reduced size.

In this work, we show a rather striking result where transition metal (TM) atoms supported on strongly interacting templates, almost maintain their atomic moments except in selected cases. Earlier experimental studies support this finding. Beckmann and Bergmann,[16] through anomalous Hall measurements, showed that a Co impurity on a Cs film has a (total) magnetic moment as large as 9 $\mu_B$, and that in bulk Cs, it has a moment of 8 $\mu_B$, while an Fe impurity has a moment of 7 $\mu_B$ both on a Cs film, and inside bulk Cs. There have been theoretical studies of TM atoms in bulk alkali metal hosts, and they also have found large moments in these systems. For Fe in Cs, using a local spin density approximation (LSDA) plus spin-orbit (SO) coupling plus on-site correlation (U) method, Kwon and Min[17] obtained magnetic moments of 3.30, 4.62 and 6.42 $\mu_B$ for LSDA, LSDA+SO and LSDA+SO+U calculations respectively, while for the same system but using an orbital-polarization correction (OPC), and the relativistic (R) LSDA, Guo[18] obtained moments of 3.46, 4.69 and 5.45 $\mu_B$ for LSDA, RLSDA, and RLSDA+OPC calculations respectively. Nearly two decades ago, Riegel et al.[19] and Kowallik et al.,[20] through magnetic susceptibility measurements, showed that Fe and Ni impurities in K, Rb and Cs hosts have large magnetic moments. Gambardella et al.[21] in their X-ray absorption spectroscopy (XAS) and X-ray magnetic circular dichroism (XMCD) measurements found total magnetic moments of 6.63, 5.59 and 3.55 $\mu_B$ respectively for Fe, Co and Ni impurities on K films. How the transition metal atoms exhibit high spin moments, and whether one can expect similar effects for molecules and larger species, however, have remained unanswered.

In this work, we have performed a theoretical study of the interaction of late transition metal atoms Cr, Mn, Fe, Co, and Ni with alkali clusters $Na_4$, $Na_5$, $Na_6$, and $Na_7$ to answer some of these questions. The results proved that the spin moment was almost intact in most cases. We show that the reason for this unexpected result lies in the fact that the electrons responsible for the chemical bonding are often distinct from the electrons responsible for the magnetic properties. Thus, the preservation of the magnetic moment can be explained through the analysis of the interaction between the *d* electrons of the TM atoms, and the *sp* electrons of the nonmagnetic host metal in the framework of the Anderson model,[22] in analogy to bulk systems with strong electron correlation.

We show that the transition metal atoms bind strongly by combining *sp*-states with *sp*-states of the $Na_n$ clusters. The *d*-states that do not interact in most cases, and interact only weakly in some selected cases, continue to carry the spin moments. Apart from the lack of mixing, the *d*-states also exhibit another interesting feature. In most cases, the energy of the unfilled *d*-states of the transition metal atom is far below the highest occupied molecular orbital (HOMO) of the separated alkali cluster. When combined, this would be expected to lead to a filling of the *d*-states through charge transfer from the higher occupied alkali metal levels. However, we find that the *d*-levels renormalize in the compound cluster, in most cases maintaining *d*- state occupations close to those in free atoms.

The rest of the paper is organized as follows. In section 2 we discuss the particular computational methods used for these studies. Section 3 presents the main results obtained for these clusters. In section 4 we analyze the results obtained, and try to gain an understanding of the electronic and magnetic properties of these clusters. Finally, section 5 presents the main conclusions of the paper.

## 2. Computational Methods

The theoretical studies were carried out within the framework of density functional theory (DFT) using the gradient corrected functionals for the exchange-correlation effects. Two complementary approaches were used to eliminate any bias towards the choice of basis functions or the numerical procedure. In one application, studies were carried out by representing the molecular orbitals by Gaussian type functions centered at the atomic sites. The actual calculations employed the deMon2k code.[23] Here, we used the PW86 generalized gradient approximation (GGA) functional,[24] and the double zeta with valence polarization (DZVP) basis sets optimized for gradient corrected exchange-correlation functionals.[25] The calculation of four-center electron repulsion integrals is avoided through the use of the variational fitting of the Coulomb potential.[26,27] Further, the auxiliary density was expanded in primitive Hermite Gaussian functions using the A2 auxiliary function set.[28] The exchange-correlation potential was calculated by a numerical integration on an adaptive grid[29,30] from the orbital density. To determine the geometry and spin multiplicity of the ground state, the configuration space was sampled by starting from several initial configurations and spin multiplicities and optimizing the geometry employing the quasi-Newton Levenberg-Marquardt method.[31,32] All structures were fully optimized in redundant coordinates without using symmetry constrains.[33] The resulting ground states were further ascertained via a frequency analysis. To eliminate any uncertainty arising from the choice of basis set or the numerical procedure, supplementary calculations were carried out for many of the clusters using plane-wave basis sets within a supercell approach as implemented in the VASP code.[34-36] Here, an energy cutoff of 500 eV was used, and the cluster was placed in a large cubic box of sides 20 Å in order to reduce its interaction with its images. The potential between the ion cores and the valence electrons were expressed in terms of projector augmented waves (PAW). The exchange-correlation effects were treated with the PW91[37] GGA functional. Brillouin zone integrations were carried out using only the $\Gamma$-point. Structures were relaxed using the conjugate gradient (CG) method for different fixed spin multiplicities, and without any symmetry constraints. We found that the all-electron cluster calculations and the plane-wave PAW supercell calculations agreed in most cases for the lowest energy spin multiplicity and structure. The only exceptions were $NiNa_5$ and $MnNa_5$ clusters. VASP predicts $NiNa_5$ to have a ground state spin multiplicity of 4, while deMon2k finds this to be 2. In case of $MnNa_5$ cluster, both the methods find a ground state spin multiplicity of 4. However, while deMon2k finds a $D_{5h}$ structure to have the lowest energy, VASP finds a $C_{4v}$ structure in that place. In what follows we report the geometries and energy values based on all-electron deMon2k calculations. All the molecular geometries were plotted with the Schakal software.[38]

## 3. Results and Discussion

**$TMNa_4$ clusters**: We start by discussing the ground state geometry, and immediate higher energy isomer (structure or multiplet) for the TM-doped Na clusters. The results are presented respectively in Figures 1 and 2. $TMNa_4$ clusters were found to have planar $C_{2v}$ or slightly distorted $C_{2v}$ structures in their ground states irrespective of the TM atom as shown in Figure 1. What is interesting is that all these clusters retain the magnetic moment found on the respective isolated TM atoms. In Table 1 we present these findings as well as the calculated gaps between the highest occupied and the lowest unoccupied molecular orbitals (HOMO-LUMO gap). The HOMO-LUMO gap is a measure of the chemical stability of the system as a large gap indicates that the cluster neither prefers to donate nor to receive charge. The HOMO-LUMO gap changes with the TM atom, varying between 0.23 eV and 0.76 eV. Also reported in Table 1 is the energy difference between the ground state of these clusters and the immediate higher energy isomer. We found that $CrNa_4$ and $NiNa_4$ are separated from their immediate higher energy isomers by ~0.1 eV. All the other clusters at this size also possess fairly close isomers. In general, the immediate higher energy isomers preserved the $C_{2v}$ symmetry and presented a smaller magnetic moment when compared to the ground state clusters (Figure 2). The only exceptions were $MnNa_4$ and $NiNa_4$ that presented less symmetric $C_s$ geometries and had the same magnetic moment.

We calculated the relative stability of different clusters via their embedding energies. Embedding energy (EE) is defined as the energy gain in incorporating a TM atom into a $Na_n$ cluster. Therefore, EE is defined by the equation

$$EE = E(Na_n) + E(TM) - E(TMNa_n), \qquad (1)$$

where $E(Na_n)$ and $E(TMNa_n)$ are the ground state total energies of the clusters, and $E(TM)$ is the total energy of the transition metal atom. According to this definition EE is positive for a bound structure, and a larger EE implies a more stable structure. In the absence of significant spin orbit interaction, the reactants and the products should have the same spin. Hence, the embedding energies were calculated enforcing spin conservation in accordance with Wigner-Witmer spin conservation rule.[39,40] Consequently, in calculating EE we used the total energy of high spin excited states of TM or $Na_n$ clusters (whichever gives the largest EE), when the ground state multiplicities did not satisfy the spin conservation rule.

**$TMNa_5$ clusters**: All the $TMNa_5$ clusters have planar $D_{5h}$ structures in their ground states, as shown in Figure 1. In the case of $NiNa_5$, VASP calculations predicted a $C_{4v}$ geometry with a magnetic moment of 3 $\mu_B$ for the ground state. However, optimizing this geometry in deMon2k led to a $D_{5h}$ structure, and resulted in higher energy than the ground state with magnetic moment of 1 $\mu_B$. As mentioned before, VASP finds the lowest energy structure for $MnNa_5$ to be $C_{4v}$. This is, however, only 0.06 eV lower in energy compared to a $D_{5h}$ structure. The HOMO-LUMO gaps and the EE's of these clusters are presented in Table I. For all TM atoms, the HOMO-LUMO gaps and EE's of the $TMNa_5$ clusters are higher than those for the corresponding $TMNa_4$ clusters. A HOMO-LUMO gap of more than 1.0 eV is generally considered high and it is interesting to note that except $NiNa_5$, all the clusters have gaps that are higher than 1.0 eV. The higher EE shows that the atoms bind more strongly to $Na_5$ than to $Na_4$ clusters. The magnetic moments on the $TMNa_5$ clusters are consistently 1 $\mu_B$ less than the moments on the corresponding $TMNa_4$ clusters. Since a $Na_5$ cluster has odd number of electrons, it seems that the unpaired electron on the $Na_5$ motif aligns antiferromagnetically to the transition metal spins. For the $TMNa_5$ series, the immediate higher in energy isomers are separated from the ground states by a relative large energy. The lowest energy difference of 0.11 eV was found for the case of $MnNa_5$, while for the rest of the clusters the energy difference was in the range of 0.3 to 0.4 eV. All the isomers presented a $C_s$ or slightly distorted $C_s$ symmetry (Figure 2), and in general the same magnetic moment when compared with the ground state clusters. Only $MnNa_5$ and $FeNa_5$ presented higher magnetic moments.

**$TMNa_6$ clusters:** This is the smallest size at which the clusters transform into three dimensional structures. Cr, Mn, and Fe doped $Na_6$ clusters presented ground state $C_{5v}$ symmetry, while $CoNa_6$ and $NiNa_6$ presented respectively $C_{3v}$ and $C_s$ symmetry as shown in Figure 1. In going from $TMNa_5$ to $TMNa_6$ clusters, the HOMO-LUMO gaps decreased in all clusters except $NiNa_6$ where it increased as shown in Table 1. In this cluster, the studies indicate a mixing between the host *sp*-states and the *d*-states of the Ni leading to a closed shell and non-magnetic ground state. In all clusters, except for NiNa$_6$, the magnetic moment increased from the TMNa$_5$ to the TMNa$_6$ series by 1 $\mu_B$, and the atomic moment of the TM atom was restored. As for the EE's, in Cr and Co doped clusters EE decreased marginally from TMNa$_5$ to TMNa$_6$ and increased for the rest of the clusters.

The energy difference between the ground state and the immediate higher energy isomers were very small for all the TMNa$_6$ clusters, and for CoNa$_6$, the two structures were nearly degenerate. While most of the higher energy isomers presented C$_{3v}$ symmetry, CoNa$_6$ and NiNa$_6$ presented C$_{5v}$ and C$_s$ symmetry respectively as shown in <span style="color:red">Figure 2</span>. All the isomers were found to have the same magnetic moment as the ground state with the exception of NiNa$_6$, in which case it increased.

**TMNa$_7$ clusters:** The ground state structures of the TMNa$_7$ clusters are shown in <span style="color:red">Figure 1</span>. MnNa$_7$ and NiNa$_7$ presented C$_s$ symmetry, while CrNa$_7$, FeNa$_7$ and CoNa$_7$ deviated from C$_s$ toward non symmetric C$_1$ geometries. In going from TMNa$_6$ to TMNa$_7$, the magnetic moments in all the clusters except NiNa$_7$ decreased as shown in <span style="color:red">Table I</span>. The moment on NiNa$_7$ became 1 $\mu_B$. This open shell structure decreases the HOMO-LUMO gap in NiNa$_7$ compared to that in NiNa$_6$. The gaps in all the other TMNa$_7$ clusters were higher than those on the corresponding TMNa$_6$ clusters. The EE's also increased in TMNa$_7$ compared to TMNa$_6$ clusters.

The immediate higher energy isomers presented C$_s$ symmetry, and CoNa$_7$ deviated from C$_s$ towards a C$_1$ geometry. Most of the isomers presented the same magnetic moment when compared with the ground states with the exceptions of MnNa$_7$ and NiNa$_7$, in which cases it increased. In CrNa$_7$, the immediate higher energy isomer was degenerate with respect to the ground state with an energy difference of only 1 meV, which is beyond the accuracy of the present DFT based methods. In FeNa$_7$ the energy difference between the immediate higher energy isomer and the ground state was ~ 0.06 eV, while the other clusters were found to be separated by relatively large energies (0.21 to 0.62 eV). So far we have reported our main findings on TMNa$_n$ clusters. In the next section we analyze these results, and try to gain some understanding of the observed phenomenon of retention of the atomic moments by the TM atoms in the TMNa$_n$ clusters.

## 4. Analysis

One of the key objectives of this work was to examine if the TM atoms maintain their atomic spin moments when combined with Na$_n$ clusters containing 4-7 atoms. These findings are summarized in Figure 3 that shows the variation of atomic moments as free Na atoms are combined with TMNa$_n$ clusters. The first thing we notice in the TMNa$_n$ clusters is that for a given TM atom, the magnetic moment shows an odd-even oscillation as one increases the number of Na atoms. The only exceptions to this rule are the Ni doped clusters. For an even number of Na atoms, we found that the TMNa$_n$ clusters presented the magnetic moment of the free TM atom, while for an odd number of Na atoms, the magnetic moment decreases by 1 $\mu_B$ when compared to the TM atom. In order to understand the origin of this behavior we analyzed the molecular orbitals (MOs) of these clusters. Additionally, knowing that when a TM atom is embedded in a bulk free electron host, a many body state where the host polarization is opposed to the TM atom can be formed, leading to the known Kondo resonance,[41] we analyzed the nature of the electron polarization on TM and on the host Na atoms through the calculation of Mulliken atomic spin charges. These are marked in Figure 1. The angular momentum character and the atomic origin of the MOs for the Na$_n$ and TMNa$_n$ clusters are presented in Figures 4 and 5 that also show the energy levels of the individual TM atoms.

In the Cr-Ni series, the atomic $d$ orbitals become progressively deeper in energy. In case of Cr and Mn, in which the $d$ orbitals are half-filled, the exchange splitting between the α and β levels is noticeably large. Moving farther along the 3$d$ row, for Fe, Co, and Ni, the exchange splitting between the α and β $d$ orbitals decreases. We found the ground state electronic configuration of the Cr and Mn atoms to be 3$d^5$4$s^1$ and 3$d^5$4$s^2$ respectively, which are the correct ground state electronic configurations as found in experiments.[42] DFT calculations within the LDA and GGAs generally tend to overestimate the stability of the $d^n s^1$ electronic configuration compared to $d^{n-1} s^2$ configuration in Fe, Co and Ni atoms. However, we employed a DZVP basis set optimized for generalized gradient functionals,[25] and were able to get the correct ground state electronic configurations of the free TM atoms. Figure 4(a) present the plot of the electronic levels of the Cr, Mn Fe, Co and Ni atoms. It should be noted that for the Fe, Co and Ni atoms, in the β spin channel, some higher energy one-electron levels are occupied, while lower energy levels are empty. This occupation ensures the correct ground state electronic configurations in an unrestricted open shell calculation, and in fact, gives lower total energy compared to the case when the lower energy one-electron β levels are occupied.

We now come back to the discussion of MOs in TMNa$_n$ clusters which are shown in panels (b) to (c) in Figures 4 and 5. In a previous publication,[40] we had reported results of a similar study on Sc, Ti and V doped sodium clusters. The main difference between the previous cases and the present is very little hybridization of the TM $d$ and Na $sp$ states. Among the CrNa$_n$ clusters, only in CrNa$_6$ there is some $spd$ hybridization as far as the occupied *d* levels are concerned. There is an overlap of empty *d* levels and anti-bonding states of the Na$_n$ host in CrNa$_5$ and CrNa$_7$. But this overlap does not affect the moments in any way.

In the MnNa$_n$ series, there is no *spd* hybridization in the α channel, while there is one occupied *spd* state in the β channel in this series. In both these series, however, there is hybridization between the TM 4*s* and Na *sp* states. In fact, this seems to be the primary binding mechanism between the Na$_n$ host and the TM dopant. The atomic moment of 6 μ$_B$ on a Cr atom is due to the 3*d$^5$* and 4*s$^1$* electrons. As a result of hybridization of the Cr 4*s* state with the Na *sp* states, part of this moment is lost. The atomic moment of 5 μ$_B$ of a Mn atom is, however, due to the half-filled 3*d* states. Due to partial hybridization of the unfilled *d* states of a Mn atom, and Na *sp* states, this moment is marginally quenched, particularly in MnNa$_5$ and MnNa$_7$. This is the reason the Cr and Mn atoms have slightly smaller-than-atomic moments in these clusters as shown through the Mulliken atomic spin charges in <span style="color:red">Figure 1</span>. What is interesting is that there is an effective ferromagnetic (anti-ferromagnetic) coupling between the localized moments on the TM atoms, and the electron cloud on the Na$_n$ host for even (odd) *n*. This is how the atomic moment of the TM is retained in CrNa$_n$ and MnNa$_n$ clusters for even *n*, and for odd *n* the moment is 1 μ$_B$ less.

Fe and Co atoms have *d$^{n-1}$s$^2$* ground state electronic configurations, 3*d$^6$*4*s$^2$* in Fe and 3*d$^7$*4*s$^2$* in Co, while for Ni the *d$^{n-1}$s$^2$* (3*d$^8$*4*s$^2$*) and *d$^n$s$^1$* (3*d$^9$*4*s$^1$*) configurations are very close in energy.[25] However, in chemical bonding these atoms present *d$^n$s$^1$* electronic configurations that give moments of 4, 3 and 2 μ$_B$ on the Fe, Co and Ni atoms respectively. <span style="color:red">Figure 4</span> and <span style="color:red">5</span> in the panels <span style="color:red">(b)</span> and <span style="color:red">(c)</span> show that there is no hybridization between Fe and Co *d* and Na *sp* states in either α or β channel at any size. Thus Fe and Co atoms retain $d^7$ and $d^8$ configuration respectively in agreement with the experimental report of Gambardella et al.[21] of Fe and Co impurities on alkali metal films.

The bonding between the $Na_n$ host and the TM atoms results primarily through the hybridization of Na *sp* and the TM 4*s* states. The moment on the TM atoms coming from the lone 4*s* electron is partly lost, and thus Fe atoms have moments between 3 and 4 $\mu_B$, and Co atoms have moments between 2 and 3 $\mu_B$ in all the Fe and Co doped clusters. As in the case of Cr and Mn doped clusters, there is an effective ferro (antiferro) magnetic coupling between the local moment on the TM atom, and the electron spin on the $Na_n$ host for even (odd) values of *n*. This odd-even oscillation gives moments of 4 and 3 $\mu_B$ on the $FeNa_n$ clusters, and 3 and 2 $\mu_B$ on the $CoNa_n$ clusters.

The Ni-doped clusters presented a slightly different picture. There are no odd-even oscillations over the size range $n = 4 - 7$. In the $NiNa_4$ cluster, there is *spd* hybridization in the $\alpha$ channel, while there is no hybridization in the $\beta$ channel. The effective *d* occupancy of the Ni atoms is 9, and there is hybridization between Na *sp* and Ni *s* states. This leaves the Ni atom with a moment of 1.14 $\mu_B$. Because of a ferromagnetic coupling between this moment and the moment on the Na host, the overall moment turns out to be 2 $\mu_B$ on the $NiNa_4$ cluster. In the $NiNa_5$ cluster, there is again *spd* hybridization in the $\alpha$ channel, but not in the $\beta$ channel. But there is hybridization between Ni 4*s* and Na *sp* states. This gives a $d^9$ configuration and a moment of ~ 1 $\mu_B$ on the Ni atoms. We find very little moment on the Na host in this case, and the net moment on the cluster is 1 $\mu_B$. There is large *spd* hybridization in $NiNa_6$ cluster. The Ni atoms is found to a have net spin moment of 0.65 $\mu_B$. In this cluster, there is an antiferromagnetic coupling between the Ni moment, and that on the Na host that compensate each other. Thus a $NiNa_6$ cluster has no net spin moment. In $NiNa_7$ there is *spd* hybridization in the $\beta$ channel, and there is also hybridization between Ni 4*s* and Na *sp* states. This gives a net moment of 0.57 $\mu_B$ on the Ni atom. There is a ferromagnteic coupling between the Ni moment and the moment on the Na host. This gives a net moment of 1 $\mu_B$ on the cluster. Thus, in NiNa$_6$ and NiNa$_7$ clusters, the behavior of odd and even number of Na atoms reverses as far as the nature of coupling of the moment on the TM atom to that on the Na host is concerned. The general finding of a $d^9$ configuration for the Ni atom in the TMNi$_n$ series resulted in agreement with Gambardella et al.[21] in their study of Ni impurities in alkali metal films.

The striking difference in the chemistry of these clusters and those of Sc, Ti and V-doped Na$_n$ clusters lies in the degree of *spd* hybridization. As we showed earlier,[40] there is a large *spd* hybridization in the TMNa$_n$ clusters for the early TM atoms and $n = 4-6$. Our preliminary results indicate that large *spd* hybridization exists up to $n = 9$[43]. In the early TM atoms, with less than half-filled 3*d* shells, *spd* hybridization increased effective *d* occupancy leading to enhancement of moments. In the late 3*d* TM doped Na$_n$ clusters, on the other hand, there is little or no *spd* hybridization. This is partly due to the fact most of the late 3*d* TM atoms have more localized (and hence smaller) *d* orbitals compared to Sc, Ti and V. Well localized *d* orbitals, and bonding through hybridization of Na *sp* and TM 4*s* states allow the TM atoms to retain their atomic moments in most of these clusters.

A comparison of the magnetic properties of these TMNa$_n$ clusters with those of TM doped Au and Ag clusters of similar sizes is also of interest here. TM atoms in small Au clusters have been found to retain their atomic character, and hence the atomic moments.[44] Torres et al.[45] reported odd-even oscillations of many physical properties of TMAu$_n^+$ clusters also. In particular, in the Cr, Mn and Fe doped clusters, the magnetic moment showed odd-even oscillations over the size range $n = 2 - 8$. Also these clusters had large magnetic moments in their ground states. In contrast, a recent work by Tono et al.[46] shows a steady decrease in the magnetic moment of $CoAg_n^-$ clusters for $n = 6 - 8$. At $n = 8$, the moment completely vanished. In this respect, the $TMNa_n$ clusters are similar to the $TMAu_n$ clusters for late TM atoms. One of the reasons for the difference between $TMAu_n^+$ and $TMNa_n$ clusters on one hand, and $CoAg_n^-$ clusters on the other, could be the structures of these clusters. In the ground state structures of $TMAu_n^+$ and $TMNa_n$ clusters, the TM is exohedrally attached to the host cluster.[45] This leads to little or no hybridization between the TM $d$ and the Na $sp$ states. In contrast, $TMAg_n^-$ clusters have ground state structures in which the TM atom is encapsulated in a Ag cage leading to greater $spd$ hybridization, and higher $d$ occupancy with increasing size. In the $CoAg_8^-$ cluster, the Co atom has a filled $3d^{10}$ configuration leading to a complete quench of the atomic moment.

**5.**

**Conclusions**

In conclusion, we have explored the electronic and magnetic properties of late $3d$ TM atom doped $Na_n$ clusters for $n = 4-7$. We found little or no $spd$ hybridization due to the fact that most of the late $3d$ TM atoms have more localized (and hence smaller) $d$ orbitals compared to the early $3d$ TM atoms. This localization of the $d$ orbitals and the bonding through hybridization of Na $sp$ and TM $4s$ states allow the TM atoms to retain their atomic moments in most of these clusters, in agreement with experimental reports of TM impurities on alkali films. In clusters with an even number of Na atoms, the atomic moment of the respective TM atoms were retained, while for odd number of Na atoms the moments were 1 $\mu_B$ less. Such odd-even oscillations can be understood in terms of a ferro (antiferro) magnetic coupling of the local moment on the TM atom, and the moment on the Na host atoms. The only exception to this rule were the $NiNa_6$ and $NiNa_7$ clusters, in which the coupling between the moment on the TM and that one the Na host were antiferromagnetic and ferromagnetic respectively, leading to ground state moments of 0 and 1 $\mu_B$ respectively on these clusters.

The HOMO-LUMO gap also presented an odd-even oscillation, where the clusters with an odd number of Na atoms showed gaps larger than 1 eV, indicating their special stability. The EE's for these clusters, generally, increased with size, indicating that these clusters will tend to bind with more Na atoms to form larger clusters. Only in case of $CrNa_5$ to $CrNa_6$, and $CoNa_5$ to $CoNa_6$ one sees marginal decrease in EEs. In order to examine if the findings on these atoms can be extended to larger sizes, we undertook the studies of a $Fe_2$ cluster interacting with $Na_n$ clusters. The results indicated that the magnetic moment of the free $Fe_2$ dimer (6 $\mu_B$) was maintained as the molecule combined with $Na_n$ (n = 1-5) clusters, and an odd even oscillation of the magnetic moment 7 (6) $\mu_B$ for odd (even) number of Na atoms appeared. This shows that these alkali clusters could indeed be thought of as molecular templates to deposit transition metal atoms and larger species without affecting the magnetic moment. As TM-doped Au and Ag clusters have been produced in experiments, we hope our work will generate interest in producing TM-doped Na or other alkali metal clusters. It will also be interesting to have a comparison between the properties of the TM-doped alkali clusters, and TM-doped coinage metal clusters.

**Acknowledgements**

J.U.R. and S.N.K. are grateful to U.S. Department of Energy Grant DE-FG02-96ER46009 for support. VASP numerical calculations for this study were carried out at the cluster computing facility in the Harish-Chandra Research Institute

(http://cluster.mri.ernet.in). Part of the deMon2k calculations were performed on the computational equipment of DGSCA UNAM, particularly at the super computer KanBalam.

**Table 1.** Electronic properties of the TMNa$_n$ clusters (TM = Cr, Mn, Fe, Co and Ni; $n$ = 4-7). Magnetic moments (μ), HOMO-LUMO gaps (Gap), and the embedding energies (EE) in the ground state, magnetic moment in the immediate higher energy isomer, and its energy difference with respect to the ground state ΔE are presented.

| Cluster | Ground State | | | Immediate higher in energy isomer | |
|---|---|---|---|---|---|
| | μ (μ$_B$) | Gap (eV) | EE (eV) | μ (μ$_B$) | ΔE (eV) |
| Na$_4$Cr | 6 | 0.59 | 1.28 | 4 | 0.101 |
| Na$_4$Mn | 5 | 0.76 | 1.06 | 5 | 0.075 |
| Na$_4$Fe | 4 | 0.68 | 1.98 | 2 | 0.075 |
| Na$_4$Co | 3 | 0.61 | 2.02 | 1 | 0.046 |
| Na$_4$Ni | 2 | 0.23 | 2.27 | 2 | 0.164 |
| Na$_5$Cr | 5 | 1.20 | 1.64 | 5 | 0.306 |
| Na$_5$Mn | 4 | 1.14 | 1.29 | 6 | 0.116 |
| Na$_5$Fe | 3 | 1.10 | 2.40 | 5 | 0.299 |
| Na$_5$Co | 2 | 1.03 | 2.51 | 2 | 0.407 |
| Na$_5$Ni | 1 | 0.39 | 2.70 | 1 | 0.379 |
| Na$_6$Cr | 6 | 0.75 | 1.63 | 6 | 0.089 |
| Na$_6$Mn | 5 | 0.74 | 1.42 | 5 | 0.029 |
| Na$_6$Fe | 4 | 0.71 | 2.44 | 4 | 0.087 |
| $Na_6Co$ | 3 | 0.61 | 2.45 | 3 | 0.003 |
| $Na_6Ni$ | 0 | 0.53 | 3.13 | 2 | 0.099 |
| $Na_7Cr$ | 5 | 1.16 | 1.81 | 5 | 0.001 |
| $Na_7Mn$ | 4 | 0.98 | 1.50 | 6 | 0.211 |
| $Na_7Fe$ | 3 | 1.08 | 2.81 | 3 | 0.056 |
| $Na_7Co$ | 2 | 0.99 | 2.97 | 2 | 0.431 |
| $Na_7Ni$ | 1 | 0.40 | 3.30 | 3 | 0.618 |

**Figure Captions**

**Figure 1. (Color online)** Ground state geometries of the $TMNa_n$ (TM = Cr, Mn, Fe, Co and Ni; n = 4-7) clusters. The bond lengths are given in Angstroms and superscripts indicate spin multiplicity. The Mulliken atomic spin charges for the symmetry inequivalent atoms are marked below them.

**Figure 2. (Color online)** Geometries of the immediate higher energy structures/multiplets isomers of the $TMNa_n$ (TM = Cr, Mn, Fe, Co and Ni; n = 4-7) clusters. The bond lengths are given in Angstroms and superscripts indicate spin multiplicity.

**Figure 3**. **(Color online)** Variation of magnetic moments ($\mu$) with $n$ in the $TMNa_n$ (TM = Cr, Mn, Fe, Co and Ni; n = 4-7) clusters.

**Figure 4.** One-electron levels (in eV) in (a) the TM atoms, (b)-(c) the $Na_n$ and the $TMNa_n$ clusters for $n$ = 4 and 5, respectively. Superscripts indicate spin multiplicities, The continuous lines represent occupied levels, and the dashed lines represent virtual levels. The atomic origin and the angular momentum character of a single level or a bunch of levels are indicated below them. Degeneracies are indicated by a number next to the level. ↑ and ↓ indicate majority and minority spin states.

**Figure 5.** One-electron levels (in eV) in (a) the TM atoms, (b)-(c) the $Na_n$ and the $TMNa_n$ clusters for $n = 6$ and 7, respectively. Superscripts indicate spin multiplicities, The continuous lines represent occupied levels, and the dashed lines represent virtual levels. The atomic origin and the angular momentum character of the levels are indicated below them. Degeneracies are indicated by a number next to the level. ↑ and ↓ indicate majority and minority spin states.

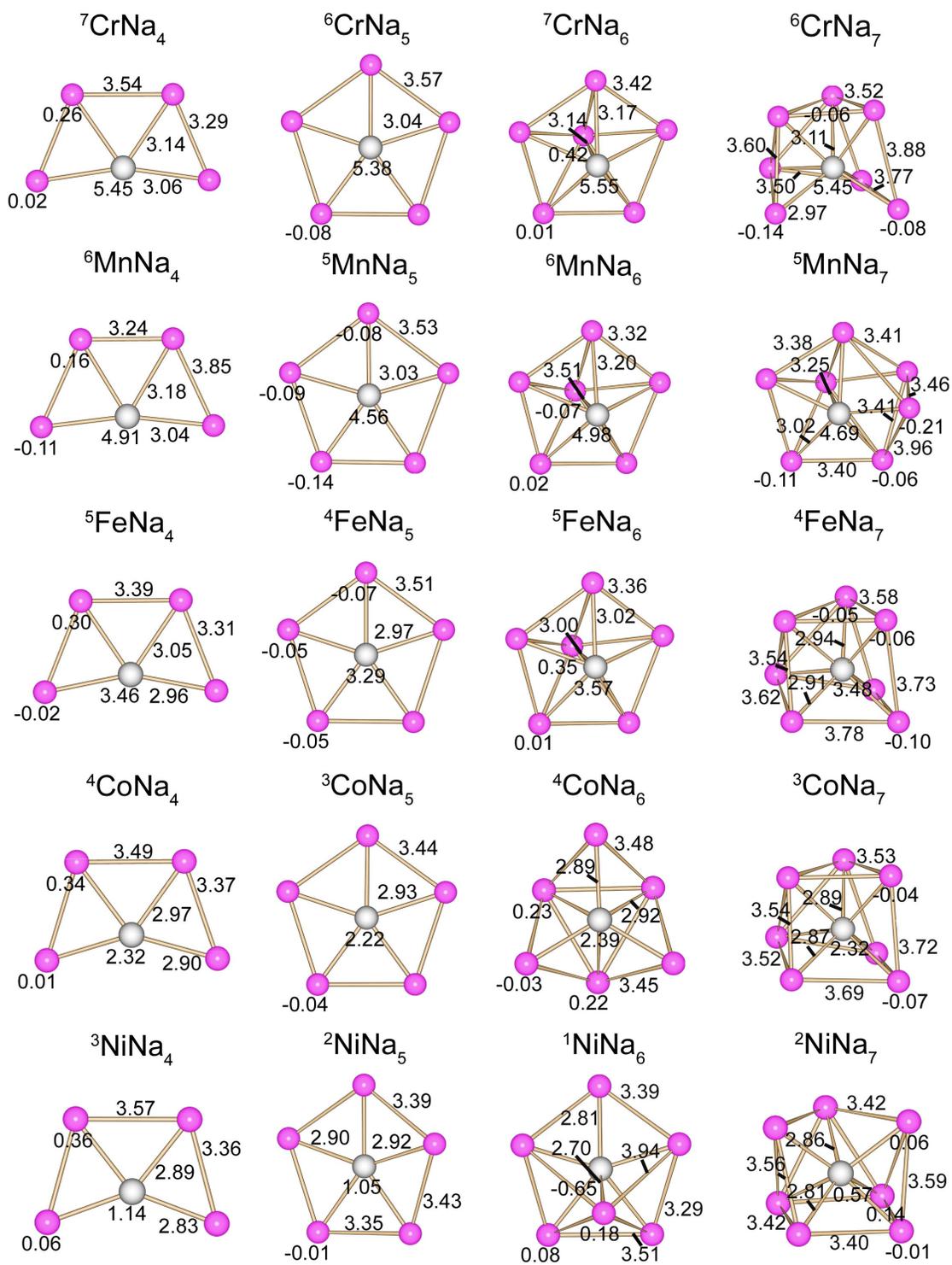

**Figure 1.**

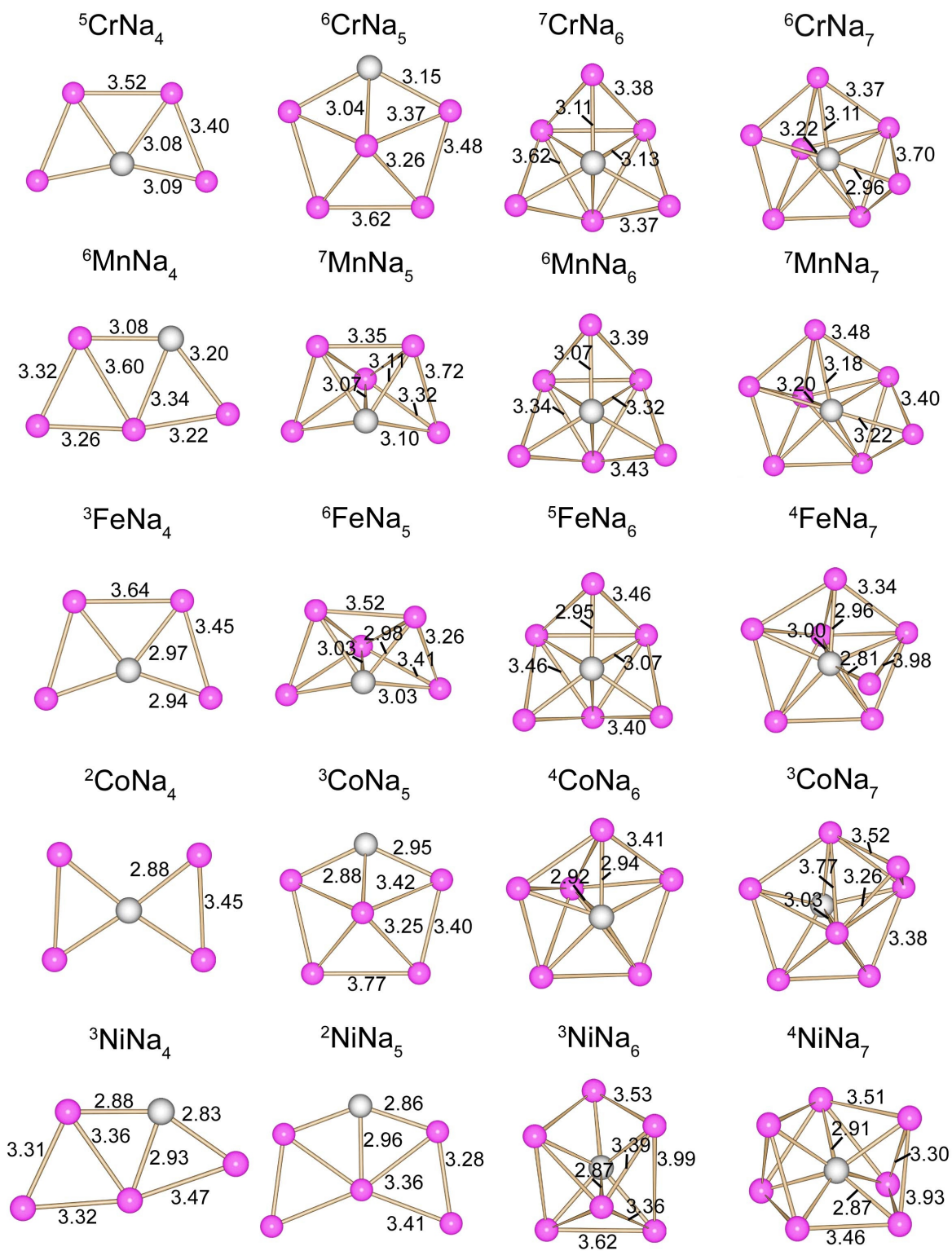

**Figure 2.**

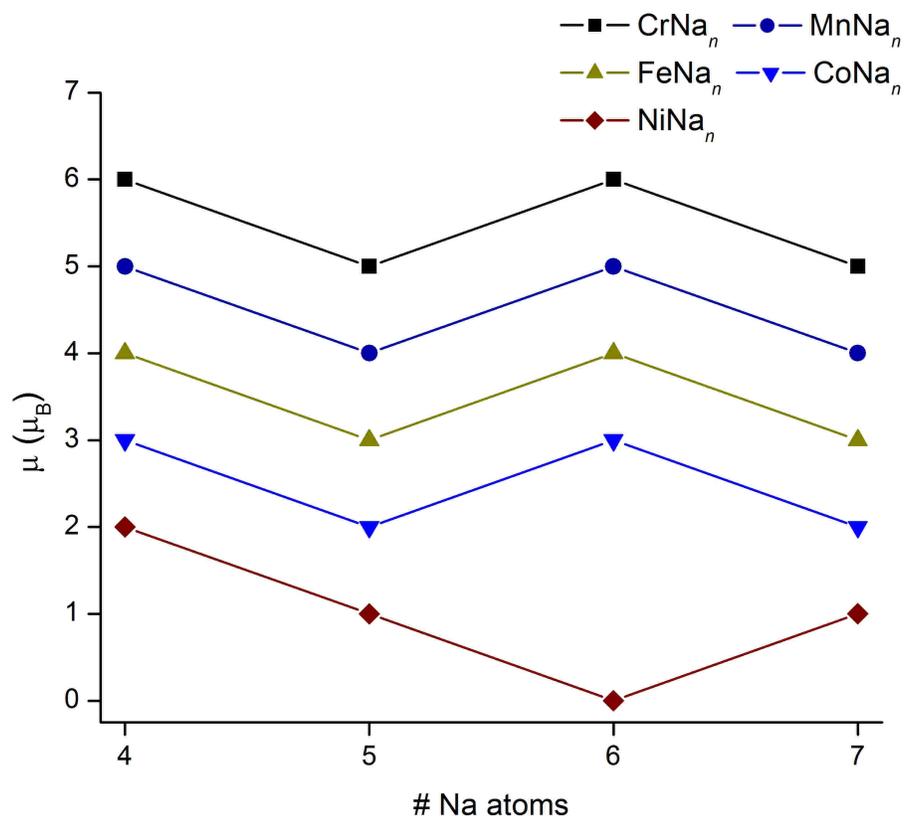

**Figure 3.**

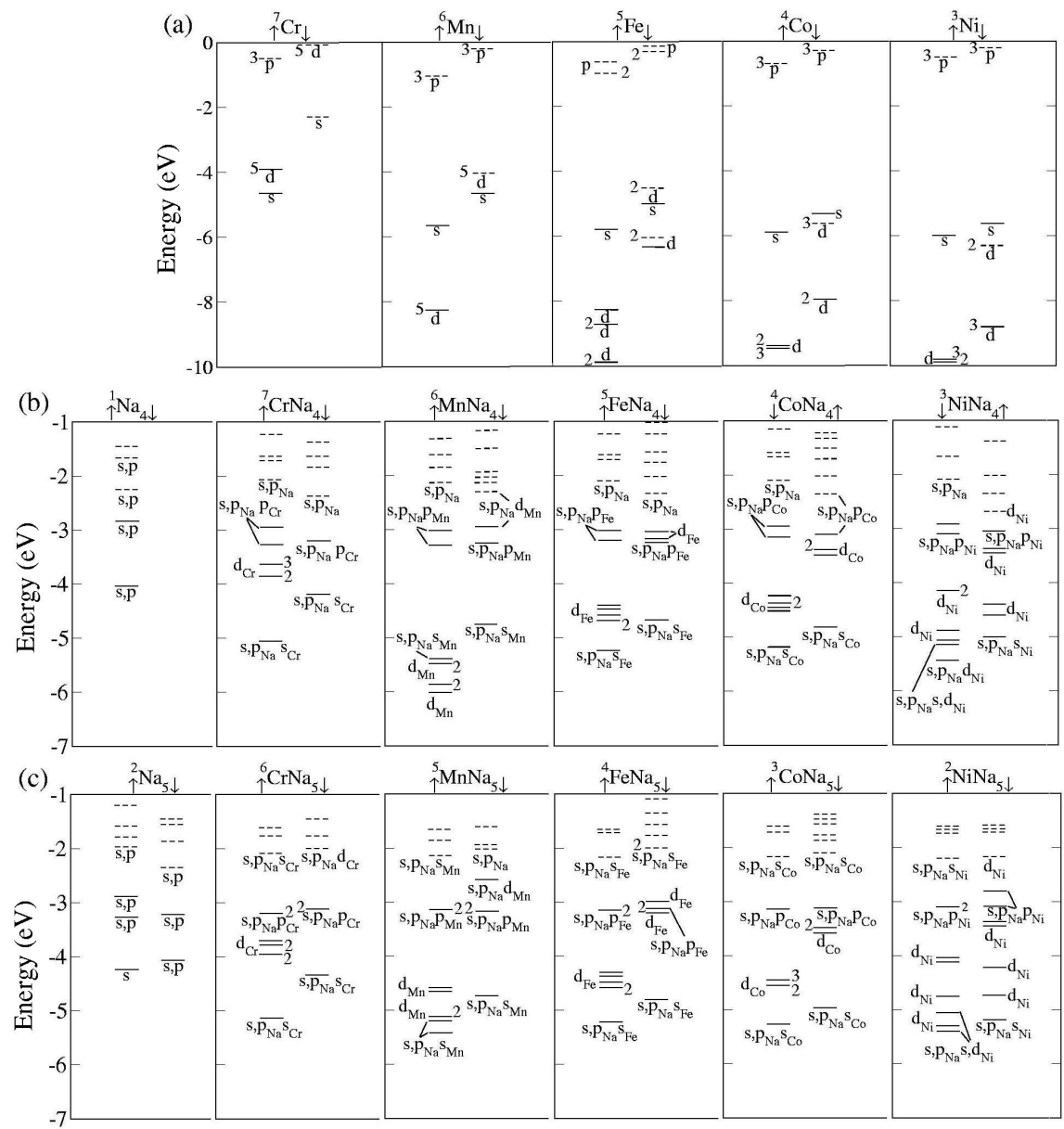

**Figure 4.**

**Figure 5.**